\documentclass[aps,prl,twocolumn,superscriptaddress,showpacs]{revtex4}
\usepackage{amsfonts}
\usepackage{amssymb}
\usepackage{amsmath}
\usepackage{graphicx}
\usepackage{dcolumn}   
\usepackage{bm}        
\usepackage{flafter}

\begin{document}


\title{Strong antiferromagnetic proximity coupling in a heterostructured superconductor Sr$_2$VO$_3$FeAs}

\author{Jong Mok Ok}
\affiliation{Center for Artificial Low Dimensional Electronic Systems, Institute for Basic Science (IBS), Pohang 37673, Korea}
\affiliation{Department of Physics, Pohang University of Science and Technology, Pohang 37673, Korea}
\author{Chang Il Kwon}
\affiliation{Center for Artificial Low Dimensional Electronic Systems, Institute for Basic Science (IBS), Pohang 37673, Korea}
\affiliation{Department of Physics, Pohang University of Science and Technology, Pohang 37673, Korea}
\author{O. E. Ayala Valenzuela}
\affiliation{Center for Artificial Low Dimensional Electronic Systems, Institute for Basic Science (IBS), Pohang 37673, Korea}
\author{Sunghun Kim}
\affiliation{Department of Physics, Korea Advanced Institute of Science and Technology, Daejeon 34141, Korea}
\author{Ross D. McDonald}
\affiliation{National High Magnetic Field Laboratory, Los Alamos National Laboratory, Los Alamos, New Mexico 87545, USA}
\author{Jeehoon Kim}
\affiliation{Center for Artificial Low Dimensional Electronic Systems, Institute for Basic Science (IBS), Pohang 37673, Korea}
\affiliation{Department of Physics, Pohang University of Science and Technology, Pohang 37673, Korea}
\author{E. S. Choi}
\affiliation{National High Magnetic Field Laboratory, Florida State University, Tallahassee, Florida 32310, USA}
\author{Woun Kang}
\affiliation{Department of Physics, Ewha Womans University, Seoul 120-750, Korea}
\author{Y. J. Jo}
\affiliation{Department of Physics, Kyungpook National University, Daegu 41566, Korea}
\author{C. Kim}
\affiliation{Department of Physics and Astronomy, Seoul National University, Seoul 08826, Korea}
\affiliation{Center for Correlated Electron Systems,Institute for Basic Science, Seoul 08826, Korea}
\author{E. G. Moon}
\affiliation{Department of Physics, Korea Advanced Institute of Science and Technology, Daejeon 34141, Korea}
\author{Y. K. Kim}
\affiliation{Department of Physics, Korea Advanced Institute of Science and Technology, Daejeon 34141, Korea}
\author{Jun Sung Kim}
\email{js.kim@postech.ac.kr}
\affiliation{Center for Artificial Low Dimensional Electronic Systems, Institute for Basic Science (IBS), Pohang 37673, Korea}
\affiliation{Department of Physics, Pohang University of Science and Technology, Pohang 37673, Korea}
\date{\today}

\begin{abstract}
We report observation of strong magnetic proximity coupling in a heterostructured superconductor Sr$_2$VO$_3$FeAs, determined by the upper critical fields $H_{c2}(T)$ measurements up to 65 T. Using the resistivity and the radio-frequency measurements for both $H \parallel ab$ and $H \parallel c$, we found a strong upward curvature of $H_{c2}^c(T)$, together with a steep increase of $H_{c2}^{ab}(T)$ near $T_c$, yielding the anisotropic factor $\gamma_H=H_{c2}^{ab}/H_{c2}^c$ up to $\sim$ 20, the largest value among iron-based superconductors.
These are attributed to the Jaccarino-Peter effect, rather than to the multiband effect, due to strong exchange interaction between itinerant Fe spins of the FeAs layers and localized V spins of Mott-insulating SrVO$_3$ layers.
These findings provide evidence for strong antiferromagnetic proximity coupling, comparable with the intralayer superexchange interaction of SrVO$_3$ layer and sufficient to induce magnetic frustration 
in Sr$_2$VO$_3$FeAs.
\end{abstract}
\smallskip

\maketitle
Heterostructures of correlated electronic systems offer novel and versatile platforms for triggering various types of interactions and stabilizing exotic electronic orders~\cite{Damascelli, Chakalian, Gozar, Satapathy, Driza,FeSeSTO1,FeSeSTO2,FeSecharge,FeAscharge}. When one of the constituent layers hosts a superconducting state, the other blocking layer in-between serves as an active spacer that controls the dimensionality and also introduces additional proximity coupling.
For example, in high-$T_c$ cuprates and iron-based superconductors (FeSCs), various types of blocking layers are used to tune the superconducting properties by changing doping levels, modifying the interlayer coupling, introducing lattice strain ~\cite{Damascelli, Chakalian, Gozar, Satapathy, Driza},
or inducing additional pairing interaction by interfacial phonons~\cite{FeSeSTO1,FeSeSTO2} or charge transfer~\cite{FeSecharge,FeAscharge}.
Particularly, when the blocking layer is magnetic, additional magnetic interactions with localized spins 
may have substantial influence on the superconducting properties, but this issue has not been much explored.

Sr$_2$VO$_3$FeAs is a naturally-assembled heterostructure and has a unique position among FeSCs.
In this compound, superconducting FeAs layers and insulating SrVO$_3$ layers are alternately stacked~\cite{Zhu,Ok} (Fig. 1a), analogous to the superlattice of FeSe/SrTiO$_3$~\cite{choi1,choi2}, but
with additional magnetic proximity coupling between Fe and V spins.
The SrVO$_3$ layers have been identified to host the Mott-insulating state~\cite{Nakamura2010b, Qian2011a,Kim2015,Ok,Ok2} in the absence of a long-range magnetism of the V spins~\cite{Ok}.
Instead, in the FeAs layers, various phase transitions occurs above the superconducting transition at $T_c \sim $ 30 K~\cite{Zhu,Ok,Cao}, including 
an intriguing $C_4$ symmetric transition at $T_{\rm HO} \sim$ 150 K~\cite{Ok,NMR,SH1,SH2} without breaking any of the underlying translational, rotational, and time reversal symmetries, reminiscent of the so-called $hidden$ order transition~\cite{YKKim}. Such a transition has never been observed in other FeSCs, and magnetic proximity coupling that induces frustration between stripe-type Fe and Neel-type V antiferromagnetism~\cite{Ok}, has been suggested to be responsible for it. However, whether or not such a magnetic proximity coupling is strong enough, and if so, whether it is ferromagnetic (FM) or antiferromagnetic (AFM), have not been clarified yet.

In this Letter, we present experimental evidences for strong AFM exchange coupling of itinerant Fe spins to localized V spins, using the upper critical field $H_{c2}$ of Sr$_2$VO$_3$FeAs single crystal for both
$H \parallel ab$ and $H \parallel c$, determined by magnetoresistance measurements up to 30 T and radio-frequency (RF) contactless measurements up to 65 T.
A strongly convex $H_{c2}^c(T)$ for $H \parallel c$ is observed in contrast to a steep linear increase of $H_{c2}^{ab}(T)$ near $T_c$ for $H \parallel ab$. In comparison with other FeSCs, we found that the Jaccarino-Peter (JP) effect with an exchange field up to $\sim$ 20 T is responsible for this unusual behavior. Our observations confirm that magnetic proximity coupling can play a critical role for inducing unusual magnetic and superconducting properties of Sr$_2$VO$_3$FeAs.

Single crystals of Sr$_2$VO$_3$FeAs were grown using self flux techniques~\cite{Ok}. The typical size of each single crystal was 200$\times$200$\times$10 $\mu m^3$. High crystallinity and stoichiometry were confirmed by X-ray diffraction and energy-dispersive spectroscopy. The single crystals show a clear superconducting transition at $T_c^{onset} \sim$ 27 K, which is somewhat lower than a maximum $T_c^{onset} \sim$ 35 K in a polycrystalline sample~\cite{Zhu}. This difference may be attributed to a partial deficiency of oxygen~\cite{SVOFA-O}. Magnetotransport measurements were carried out using conventional six-probe method in a 14 T Physical Property Measurement System and a 33 T Bitter magnet at the National High Magnetic Field Lab., Tallahassee. RF contactless measurements up to 65 T were performed in the National High Magnetic Field Lab., Los Alamos.

\begin{figure}[t]
\centering
\includegraphics[width=8.6cm]{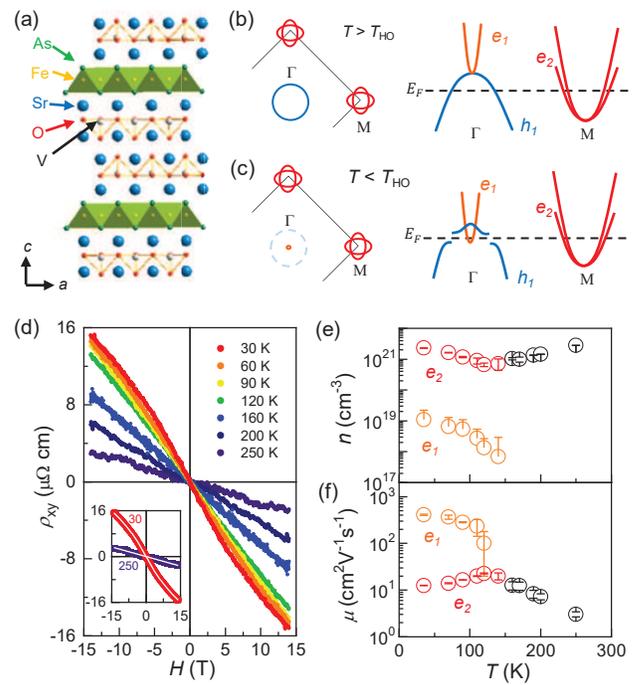}
\caption{\label{Fig1}(Color online)
(a) Crystal structure of Sr$_2$VO$_3$FeAs consisting of FeAs layers and SrVO$_3$ layers. (b, c) Schematic illustration of Fermi surface and band structures above (b) and below (c) $T_{\rm HO}$ = 150 K. (d) Magnetic field dependence of the Hall resistivity $\rho_{xy}(H)$ measured at different temperatures. The solid lines in the inset represent best fits for Hall data. (e, f) Temperature dependence of the carrier density (e) and the mobility (f), extracted from the fits using one band model (black) for $T > T_{\rm HO}$ and two-band model (red and orange) for $T < T_{\rm HO}$.}
\end{figure}

Before discussing the upper critical field $H_{c2}$ of Sr$_2$VO$_3$FeAs, we first consider the Fermi surface reconstruction across the $C_4$ symmetric transition $T_{\rm HO}$ $\approx$ 150 K. According to recent ARPES results on Sr$_2$VO$_3$FeAs in the wide range of temperature, the heavy hole FS centered at the $\Gamma$ point of Brillouin Zone (BZ), denoted $h_1$ in Figs. 1b and 1c, has a relatively strong $k_z$ dispersion and becomes fully gapped below $T_{\rm HO}$. In contrast, the two dimensional electron FS at the $M$ point ($e_2$ in Figs. 1b and 1c) remains gapless. Concomitantly the additional small electron FS ($e_1$ in Figs. 1b and 1c), which is absent in the calculated band structures of Sr$_2$VO$_3$FeAs~\cite{YKKim}, is introduced at the $\Gamma$ point, as illustrated in Figs. 1b and 1c. Because of this unusual band selective gap opening at $T_{\rm HO}$, low-energy electronic structures of Sr$_2$VO$_3$FeAs are significantly reconstructed to yield two separate electron FSs ($e_1$ and $e_2$) with strong mismatch in size(Supplementary Fig. S1)~\cite{supp}. These features are highly distinct from those of other FeSCs.

The FS reconstruction of Sr$_2$VO$_3$FeAs is also probed by the field dependent Hall resistivity $\rho_{xy}(H)$ of Sr$_2$VO$_3$FeAs at different temperatures under magnetic field up to 14 T (Fig. 1d). Above $T_{\rm HO} \sim 150$ K, a linear field dependence of $\rho_{xy}(H)$ with a negative slope is observed up to $H$ = 14 T, similar to the cases of other FeSCs, in which charge conduction is dominated by electron FSs with a high mobility~\cite{hall1}. The contribution of the hole FSs usually appear in $\rho_{xy}(H)$ at low temperatures with a positive slope~\cite{FeSehall,ba122hall,Hall112, Hall111, Hall122, Hall1111,supp}, but is completely absent in Sr$_2$VO$_3$FeAs.
Instead we found that a non-linear field dependence in $\rho_{xy}(H)$ suddenly appears below $T_0$, which is well reproduced by the two-band model with two distinct electron carriers. Using a constraint of $1/\rho_{xx}(T)=\sum n_i e \mu_i$, the fit to the two-band model gives us the temperature dependent carrier density ($n_i$) and carrier mobility ($\mu_i$) as shown in Figs. 1e and 1f.
Clearly, additional electron carriers ($e_1$) with lower density but a higher mobility are induced on top of the high density electron carriers ($e_2$).
The densities of the two electron carriers are estimated to be $\approx1.1\times10^{19}$ cm$^{-3}$ and $\approx$ 2.3$\times10^{21}$ cm$^{-3}$, which are in good agreement with those of the $e_1$ FS at the $\Gamma$ ($\approx$ $2.0\times10^{19}$ cm$^{-3}$) and the $e_2$ FS at the $X$ point ($\approx$ $1.1\times10^{21}$), obtained by recent ARPES studies~\cite{Kim2015}.
This additional conduction channel of the small FS ($e_1$) with high mobility compensate for the loss of conduction from the gapped hole FS below $T_{\rm HO}$, which may explain a weak resistivity anomaly across $T_{\rm HO}$.

\begin{figure*}
\begin{center}
\includegraphics[width=17.2cm]{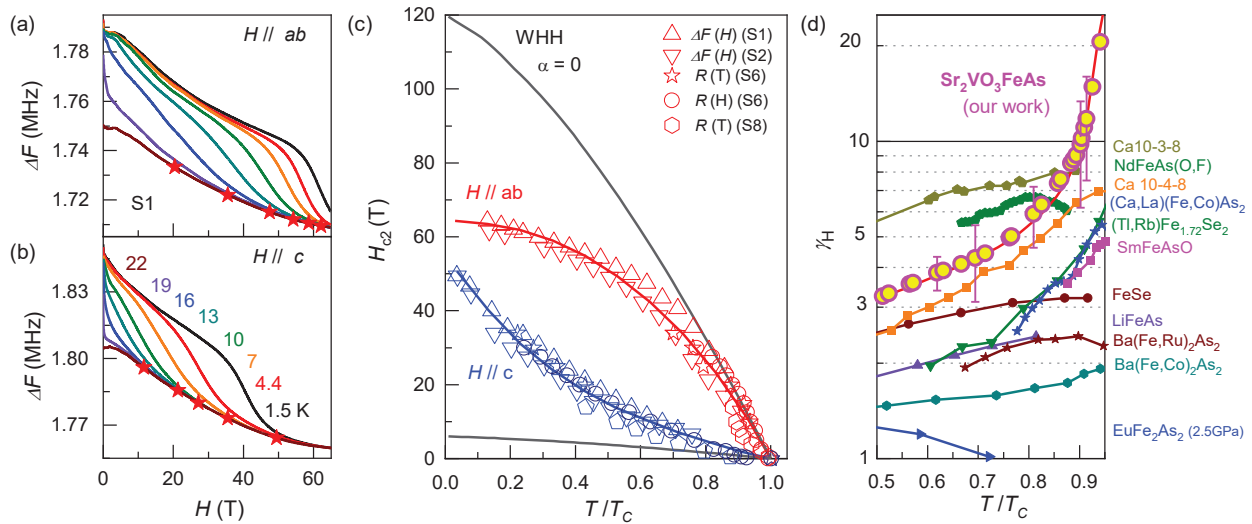}
\caption{\label{fig1}Magnetic field dependence of the radio-frequency for (a) $H \parallel  ab$ and (b) $H \parallel c$. Red stars show estimated $H_{c2}$. (c) Upper critical field $H_{c2}(T)$ for $H \parallel ab$ (red open symbols) and $H \parallel c$ (blue open symbols) as a function of the normalized temperature $T/T_c$ for the four crystals (S1, S2, S6, S8) estimated from the TDO and resistivity measurements. Red and blue solid lines are Jaccarino-Peter fits for $H \parallel  ab$ and $H \parallel c$, respectively, and the black solid lines are Werthamer-Helfand-Hohenberg curves with $\alpha = 0$ for comparison. (d) Temperature dependent anisotropic factor $\gamma_H$ of Sr$_2$VO$_3$FeAs and other FeSCs~\cite{D122_3, g1038, gNd111, g1048, gCa112, gRbFeSe2, gSm1111, gFeSe, gLi111, gBaRuCo122, gEu122}. Red solid line shows $\gamma_H$ of Sr$_2$VO$_3$FeAs calculated from the $H_{c2}(T)$ fits for comparison. }
\end{center}
\end{figure*}

Now we focus on the upper critical field $H_{c2}$ of Sr$_2$VO$_3$FeAs single crystals, obtained from RF measurements and the resistivity (Fig. 2). The radio-frequency curves as a function of magnetic fields along $H \parallel ab$ and $H \parallel c$ yield $H_{c2}(T)$ at various temperatures (Figs. 2a and 2b). Here we determined $H_{c2}$ by taking the magnetic field at which the steepest slope of the radio-frequency intercepts the normal-state background.
Temperature and magnetic field dependence of resistivity $\rho_{xx}$ were also used to determine $H_{c2}(T)$ under magnetic field up to 33 T (Supplementary Fig. S2)~\cite{supp} . Using the criterion of 50$\%$ of resistive transition, we obtained $H_{c2}(T)$, consistent with that from the RF contactless measurements.
We note that using different criteria for $H_{c2}$ in the RF and the resistivity measurements obtained qualitatively the same $H_{c2}(T)$ behaviour(Supplementary Fig. S3)~\cite{supp}.

Figure 2 (c) shows $H_{c2}(T)$ curves as   function of the normalized temperature ($t=T/T_c$) for $H \parallel ab$ and $H \parallel c$. We found that $H_{c2}(T)$ curves taken from different samples and different measurements are consistent with each other.
Depending on the magnetic field orientations, $H_{c2}(T)$ exhibits different behaviors. For $H \parallel ab$, $H_{c2}^{ab}(T)$ shows a concave temperature dependence with saturation at low temperatures. This shape is typically observed in many FeSCs~\cite{hc2,hc2review} in which the Pauli limiting effect dominates over other pair-breaking mechanisms. In contrast, $H_{c2}^c(T)$ for $H \parallel c$ shows a strongly convex behaviour with a strong upward curvature. The similar convex behaviours of $H_{c2}(T)$ have been rarely observed, except in some FeSCs including Ba(Fe,Co)$_2$As$_2$~\cite{D122_3}, (Sr,Eu)(Fe,Co)$_2$As$_2$~\cite{D122_4}, LaFeAs(O,F)~\cite{D1111_1}, and NdFeAs(O,F)~\cite{gNd111}. However their upward curvature of $H_{c2}(T)$ is far less significant than found in Sr$_2$VO$_2$FeAs.

This strong anisotropic behavior of $H_{c2}$ in Sr$_2$VO$_3$FeAs can be quantified by the anisotropy factor $\gamma_H=H^{ab}_{c2}/H^{c}_{c2}$. We plot the temperature dependent $\gamma_H$ for Sr$_2$VO$_3$FeAs together with other FeSCs in Fig. 2(d). Near $T_c$, the slope of $H_{c2}(T)$ is estimated to be $dH_{c2}/dT|_{T_c}\simeq$ -7.4 T/K for $H \parallel ab$ and $\simeq$ -0.2 T/K for $H \parallel c$, in Sr$_2$VO$_3$FeAs, resulting in $\gamma_H \sim$ 20 at $T \approx T_c$. This is the highest $\gamma_H$ found in FeSCs. As shown in Fig. 2(d), the typical values of $\gamma_H$ are $\simeq$ 2-3 in the so-called 122 compounds and $\gamma_H\simeq$ 5-6 in the 1111 compounds.
Usually, the thicker blocking layer between the superconducting layers induces the stronger anisotropy of $H_{c2}$ with a larger $\gamma_H$.
The $\gamma_H$ values of various FeSCs with a different thickness ($d$) of the blocking layer follow an empirical relation $\gamma_H$/$d$ $\sim$ 0.65 $\rm \AA^{-1}$ (Supplementary Fig. S4)~\cite{supp}. However, Sr$_2$VO$_3$FeAs has $\gamma_H$ $\approx$ 20, which is  by a factor of two larger than what is expected. 
This observation indicates that the relatively thick blocking layer in Sr$_2$VO$_3$FeAs alone cannot explain the observed $\gamma_H$ and also its strong temperature dependence.

\begin{figure}
\begin{center}
\includegraphics[width=8.6cm]{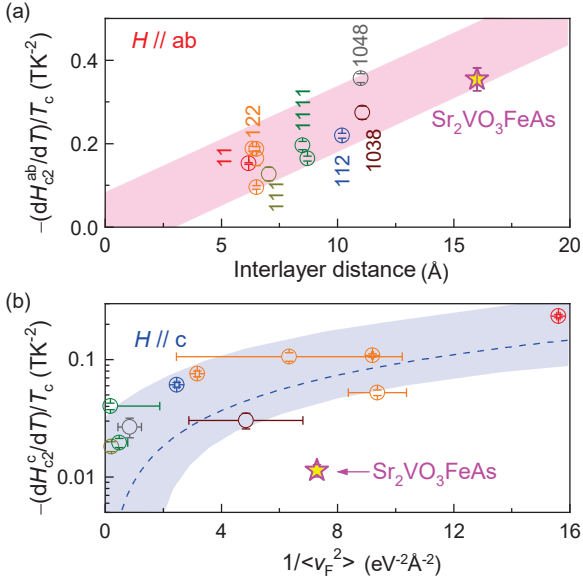}
\caption{\label{fig2}(Color online) (a) Interlayer distance dependence of normalized slope of $H_{c2}(T)$ near $T_c$ for $H \parallel ab$. The normalized slope of $H_{c2}(T)$ increases in proportion to the interlayer distance. (b) Normalized slope of $H_{c2}(T)$ near $T_c$ for $H \parallel c$ as a function of $<v_F^2>$. Sr$_2$VO$_3$FeAs is clearly out of the trend of other FeSCs~\cite{A11, A122_1, A122_2, A122_3, A111, A1111_1, A1111_2, A1111_3, A112_1, Kim2015}.}
\end{center}
\end{figure}

In comparison with other FeSCs, we found that the exceptionally small $dH_{c2}^{c}/dT$ near $T_c$ for $H\parallel c$ is crucial to the large $\gamma_H$ in Sr$_2$VO$_3$FeAs. In the case of $H\parallel ab$, the normalized slope of the upper critical field
at $T_c$, $-(dH_{c2}^{ab}/dT)/T_c$ is closely related to the diffusivity along the $c$-axis and thus is sensitive to the interlayer distance. 
Sr$_2$VO$_3$FeAs nicely follows the linear trend of $-(dH_{c2}^{ab}/dT)/T_c$ as a function of the thickness of the blocking layer $d$ (Fig. 3a). 
The distinct behaviour of Sr$_2$VO$_3$FeAs is observed for $H\parallel c$. In the case of $H\parallel c$, $-(dH_{c2}^{c}/dT)/T_c$ is more sensitive to the electronic structure of the FeAs layer than to the interlayer distance.
In conventional superconductors, $-(dH_{c2}^{c}/dT)/T_c$ is known to be proportional to $1/\langle v_F^2\rangle$ (Fig. 3b, blue dotted-line). The strong correlation between $-(dH_{c2}^{c}/dT)/T_c$ and $1/\langle v_F^2\rangle$ is confirmed in FeSCs (Fig. 3b). The data for Sr$_2$VO$_3$FeAs, however, clearly deviate from this trend and show the lowest $-(dH_{c2}^{c}/dT)/T_c$ value, leading to the largest $\gamma_H$ among the FeSCs.

For many FeSCs, temperature dependent $H_{c2}^{c}(T)$ has been understood using the two-band dirty-limit model~\cite{multihc2}. In this model, the intra- and inter-band coupling ($\lambda_{11,22}$ and $\lambda_{12,21}$) and diffusivity of each band ($D_1$, $D_2$) determines $H_{c2}^{c}(T)$ (See Supplementary Fig. S5)~\cite{supp}. The two-band model can also reproduce the strongly convex behavior of $H_{c2}^c(T)$ of Sr$_2$VO$_3$FeAs, if we assume
dominant interband coupling ($\lambda_{11}\lambda_{22}<\lambda_{12}\lambda_{21}$)
and an unusually large $\eta=D_1^{ab}/D_2^{ab} \sim 30$ (Supplementary Figs. S5 and S6)~\cite{supp}. 
We note however that most of the FeSCs show 
a concave $H_{c2}^c(T)$, and even in a few cases, like Ba(Fe,Co)$_2$As$_2$~\cite{D122_3}, LaFeAs(O,F) ~\cite{D1111_1} or NdFeAs(O,F)~\cite{gNd111}, that show a convex $H_{c2}^c(T)$, the highest estimated $\eta$ is $\sim$ 10~\cite{hc2review}, which is far less than the estimate $\eta \sim$ 30 for Sr$_2$VO$_3$FeAs.
Furthermore, the hole FS ($h_1$) centered at the $\Gamma$ point of BZ is gapped out below $T_{\rm HO}$ (Figs. 1b and 1c), and therefore cannot participate in the interband superconducting pairing. The remaining interband coupling channel is between electron FSs ($e_1$ and $e_2$) centered at $\Gamma$ and $M$ points (Fig. 1c). However, considering their drastic size difference by two orders of magnitude, confirmed by ARPES and Hall resistivity results, they are unlikely to produce strong interband coupling. These observations suggest that the conventional multiband effect cannot be the origin of the observed $H_{c2}^c(T)$ of Sr$_2$VO$_3$FeAs.

\begin{figure}
\begin{center}
\includegraphics*[width=8.6cm]{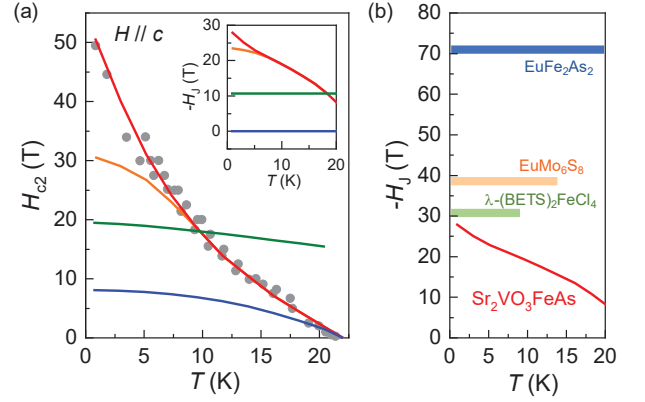}
\caption{\label{fig2}(Color online)
(a) Temperature dependent upper critical field $H_{c2}(T)$ of Sr$_2$VO$_3$FeAs for $H \parallel c$. $H_{c2}^c(T)$ data (grey symbols) and Jaccarino-Peter fits by using different temperature dependence of exchange fields $H_J$ as shown in the inset (solid line). The best fit of $H_{c2}^c(T)$ (red solid line) was obtained using an unsaturating $H_J$ at low temperature, whereas saturating $H_J$ curves produce concave $H_{c2}^c(T)$ (orange solid lines). (b) The estimated exchange field $H_J$ for Sr$_2$VO$_3$FeAs (this work), EuFe$_2$As$_2$ (Ref.~\onlinecite{Eu122}), EuMo$_6$S$_8$ (Ref.~\onlinecite{EuMoS1,EuMoS2}), $\lambda$-(BETS)$_2$FeCl$_4$ (Ref.~\onlinecite{FeCl4_1,FeCl4_2}).
}
\end{center}
\end{figure}

Instead magnetic coupling between itinerant Fe and localized V spins can offer a natural explanation for 
a strongly convex behaviour of $H_{c2}^c(T)$. Recent high field magnetoresistance (MR) results reveal a strong negative MR with a clear kink at $\sim$ 38 T for $H \parallel c$, in contrast to the monotonic positive MR for $H \parallel ab$ ~\cite{YKKim}. These results resemble the case of EuFe$_2$As$_2$~\cite{Eu122} and indicate a field-induced saturation of magnetic V moment for $H \parallel c$ but not for $H \parallel ab$. Strong exchange coupling $J$ of itinerant Fe electrons to localized V spins is then expected to introduce a net internal magnetic field $H_J$ = $J\left<S\right>/g_m\mu_{\rm B}$, which is referred to as the JP effect~\cite{jacc}.
With AFM exchange interaction ($J <$0), a negative $H_J$ is produced by polarization of V spins along the external field, particularly for $H \parallel c$. 
For paramagnetic V spins, their susceptibility and thus $H_J$ increase with lowering temperatures. Therefore, $H_J$ compensates for the external field and enhances $H_{c2}(T)$ at low temperature with large external fields.  
This trend results in a convex $H_{c2}(T)$, as observed in Fig. 2(c).

In the JP model with multiple pair-breaking including the exchange field due to the localized moments~\cite{jacc}, $H_{c2}(T)$ can be described as
\begin{small}
\begin{align}\notag
ln\frac{1}{t}&=\left(\frac{1}{2}+\frac{i\lambda_{SO}}{4\gamma}\right)\times\Psi\left(\frac{1}{2}+\frac{h+i\lambda_{SO}/2+i\gamma}{2t}\right) \\
&+\left(\frac{1}{2}-\frac{i\lambda_{SO}}{4\gamma}\right)\times\Psi\left(\frac{1}{2}+\frac{h+i\lambda_{SO}/2-i\gamma}{2t}\right)-\Psi\left(\frac{1}{2}\right),
\end{align}
\end{small}
where $\gamma=[\alpha^2(h+h_J)^2-\lambda_{SO}^2]^{\frac{1}{2}}$,  $t=T/T_c$, $h= 0.281H_{c2}/H_{c2}^*$, $h_J=0.281H_J/H_{c2}^*$, $H_{c2}^*$ is orbital critical field at $T$= 0 K, $\Psi$ is the digamma function, $\lambda_{SO}$ is spin-orbit scattering parameter and $\alpha$ is the Maki parameter. Using $\lambda_{SO}$ = 0.3 and $\alpha$ = 2.1, we successfully reproduced $H_{c2}(T)$ for $H$ $\parallel$ $c$ (Figs. 2c and 4a). We note that in the JP model, neither a temperature independent $H_J$ (green solid line) nor a saturating $H_J$ at low temperature (orange solid line) reproduces the observed upward curvature in $H_{c2}^c(T)$. 
The increasing $H_J$ at low temperatures (red solid line) is found to be crucial to reproduce the convex $H_{c2}(T)$, which is consistent with the paramagnetic state of V spins~\cite{Ok}. 
The best fit for $H$ $\parallel$ $c$ yields the negative exchange field $H_J$ increasing in magnitude up to $\sim$ 30 T with lowering temperature (red solid line in the inset of Fig. 4a). We also found that if the maximum $H_J$ is lower than $\sim$ 10 T, the calculated $H_{c2}(T)$ becomes similar to the WHH prediction. For $H$ $\parallel$ $ab$, therefore, the WHH-like behaviour can be explained by the upper bound of $H_J$ $\sim$ 10 T. This anisotropic $H_J$ may be due to the magnetic anisotropy of V spins, which is consistent with the anisotropic MR showing the conventional positive MR for $H \parallel {ab}$ and the negative MR for $H \parallel c$ below $T_{HO}$ ~\cite{YKKim}. 


The maximum $H_J$ of Sr$_2$VO$_3$FeAs, estimated for $H \parallel c$, is comparable with those of other superconductors that show the JP effect, including EuMo$_6$S$_8$~\cite{EuMoS1,EuMoS2} and $\lambda$-(BETS)FeCl$_4$~\cite{FeCl4_1,FeCl4_2}, and far less than  $H_J$ $\sim$ 75 T of EuFe$_2$As$_2$ (Fig. 4b). However, considering the smaller $\left<S\right>$ = 1 of V spins with 3$d^2$ configurations than $\left<S\right>$ = 7/2 of Eu spins, the coupling constant $J$ $\sim$ 2.3 meV is comparable in EuFe$_2$As$_2$ and Sr$_2$VO$_3$FeAs, suggesting that they share the AFM interlayer exchange interaction. Despite the similarity, magnetism of Sr$_2$VO$_3$FeAs is highly distinct from that of EuFe$_2$As$_2$. In EuFe$_2$As$_2$, Eu magnetism is induced by RKKY interaction due to itinerant Fe electrons 
~\cite{Zapf}. In contrast, the SrVO$_3$ layers in Sr$_2$VO$_3$FeAs have their own superexchange interaction ($J_{\rm S}$), competing with the RKKY interaction through the FeAs layers. The proximity coupling strength $J$ $\sim$ 2.3 meV is comparable with the superexchange interaction of V spins, $J_{\rm S}$ $\sim$ 1.6 meV, estimated from the Curie-Weiss temperature $\Theta_{\rm CW}\sim 100$ K~\cite{Ok}. 
Furthermore, it is AFM type, in contrast to the FM type expected in total energy calculations~\cite{Ok, Mazin}.
Such a significant AFM proximity coupling is effective to frustrate two distinct magnetic instabilities, Neel-type in V spins and stripe-type in Fe spins in Sr$_2$VO$_3$FeAs. This magnetic frustration has drastic effect on magnetism by destabilizing these conventional AFM orders in FeAs and SrVO$_3$ layers~\cite{Ok}, which precipitates a $C_4$ symmetric transition without long-range magnetic order.

In conclusion, based on the upper critical field $H_{c2}$ and Hall resistivity $\rho_{xy}$ results, we show that strong convex behavior of $H_{c2}(T)$ and the highly anisotropic $H_{c2}$ with the largest $\gamma_H \sim$ 20 among the FeSCs are due to magnetic proximity coupling with the neighboring localized V spins by the JP effect, rather than by the multi-band effect.
These findings demonstrate that a heterostructured Sr$_2$VO$_3$FeAs is a unique example in which the exotic electronic order, triggered by the magnetic proximity coupling, significantly affects the low energy electronic structure and superconductivity. Our work highlights that correlated heterostructures with FeSCs offer novel and fertile grounds for studying the interplay between superconductivity and the hidden competing orders.

The authors thank Y. G. Bang for fruitful discussion. We also thank H. G. Kim in Pohang Accelerator Laboratory (PAL) for the technical support. This work was supported by the Institute for Basic Science (IBS) through the Center for
Artificial Low Dimensional Electronic Systems (no. IBS-R014-D1) and by the National Research Foundation of Korea (NRF) through SRC (Grant No. 2018R1A5A6075964) and the Max Planck-POSTECH Center for Complex Phase Materials (Grant No. 2016K1A4A4A01922028). W.K. acknowledges the support by NRF (No. 2018R1D1A1B07050087, 2018R1A6A1A03025340), and Y.J.J. was supported by NRF (No. NRF-2019R1A2C1089017). A portion of this work was performed at the National High Magnetic Field Laboratory, which is supported by the National Science Foundation Cooperative Agreement No. DMR-1644779 and the state of Florida.

\end{document}